\documentclass[aps,prl,showpacs,twocolumn,floatfix]{revtex4}
\usepackage[dvips]{graphicx}
\DeclareGraphicsExtensions{.eps}
\usepackage{amsmath}
\usepackage{amsfonts}
\usepackage{amssymb}
\newcommand{\bel}[1]{\begin{equation}\label{#1}}

\newcommand{\be}{\begin{equation}}
\newcommand{\qe}{\end{equation}}
\newcommand{\ba}{\begin{eqnarray}}
\newcommand{\ea}{\end{eqnarray}}
\newcommand{\rf}[1]{(\ref{#1})}
\def\mywidth{\columnwidth}
\newcommand{\fig}[1]{\includegraphics[width=0.95\columnwidth]{#1}}

\begin{document}

\title{Delays, connection topology, and synchronization of coupled chaotic maps}
\date{\emph{Physical Review Letters} {\bf 92}, 144101 (2004). DOI: 10.1103/PhysRevLett.92.144101}

\author{Fatihcan M. Atay}
\email{atay@member.ams.org}
\homepage{http://personal-homepages.mis.mpg.de/fatay}
\author{J\"urgen Jost}
\email{jjost@mis.mpg.de}
\altaffiliation{Also at Santa Fe Institute, 1399 Hyde Park Road, Santa Fe,
NM 87501, USA, jost@santafe.edu}
\affiliation{Max Planck Institute for Mathematics in the Sciences,
Leipzig 04103, Germany}
\author{Andreas Wende}
\affiliation{Department of Computer Science,
University of Leipzig, Leipzig 04109, Germany}
\pacs{05.45.Ra, 05.45.Xt, 89.75.-k  \hfill \textbf{Copyright:} \emph{American Physical Society} }%

\begin{abstract}%
We consider networks of coupled maps where the connections between
units involve time delays. We show that, similar to the undelayed case, 
the synchronization of the network depends on
the connection topology,
characterized by the spectrum of the graph Laplacian.
Consequently, scale-free and random networks are capable of synchronizing despite
the delayed flow of information, whereas regular networks with 
nearest-neighbor connections and their small-world variants generally exhibit poor
synchronization. On the other hand, connection delays can actually be conducive
to synchronization, so that it is possible for the delayed system to
synchronize where the undelayed system does not.
Furthermore, the delays determine the synchronized dynamics,
leading to the emergence of a wide range of new collective behavior
which the individual units are incapable of producing in isolation. 
(\href{http://link.aps.org/abstract/PRL/v92/e144101}{http://link.aps.org/abstract/PRL/v92/e144101})

\vspace{0.2cm}

\end{abstract}%

\maketitle

Recent years have witnessed a growing interest in the dynamics of
interacting units. Particularly, a large number of studies have been devoted
to synchronization in a variety of systems (see \cite{Pikovsky-book01} and the
references therein), including the coupled map lattices introduced by Kaneko
\cite{Kaneko84}. Usually, such systems have been investigated under the
assumption of a certain regularity in the connection topology, where units are
coupled to their nearest neighbors or to all other units. Lately, more general
networks with random, small-world, scale-free, and hierarchical architectures have been
emphasized as appropriate models of interaction
\cite{Gade95,Gade96,Watts98,Barabasi99,Cosenza01}. 
On the other hand, realistic modeling of many large networks 
with non-local interaction inevitably requires connection delays to be taken
into account,
since they naturally arise as a consequence of finite information 
transmission and processing speeds among the units. 
Some numerical studies have regarded synchronization under 
delays for special cases such as globally coupled logistic maps
\cite{Jiang00} or carefully chosen delays \cite{Masoller03}.
In this Letter we consider synchronization of coupled chaotic maps 
for general network architectures
and connection delays.
Because of the presence of the delays, the constituent units are unaware 
of the present state of the others; 
so it is not evident \emph{a priori} that such a collection of 
chaotic units can operate in unison, i.~e.~synchronize. 
Based on analytical calculations, we show that this is indeed 
possible, and in fact may be facilitated by the presence of delays. 
Moreover, while the connection topology is important for synchronization, 
the delays have a crucial role in determining the resulting collective dynamics.
As a result, the synchronized system can exhibit a plethora of new behavior
in the presence of delays.
We illustrate the results by numerical simulation of large networks of 
logistic maps.

We consider a finite connected graph $\Gamma$ with nodes (vertices) $i$, 
writing $i \sim j$ when $i$ and $j$ are neighbors, that is, connected by an edge, 
and with the number of neighbors of $i$ denoted by $n_i$. On $\Gamma$, we have
a dynamical system with discrete time $t \in \mathbf{Z}$, with the
state $x_i$ of $i$ evolving
according to 
\begin{equation}
x_{i}(t+1)=f(x_{i}(t))+\varepsilon\left(\frac{1}{n_{i}}\sum_{\substack{j\\j\sim
i}}  f(x_{j}(t-\tau))-f(x_{i}(t))\right)  .\label{coupled}%
\end{equation}
Here $f$ is a differentiable function mapping some finite interval, 
say $[0,1]$, to itself, $\varepsilon \in [0,1]$ is the coupling strength, 
and $\tau \in \mathbf{Z^+}$ is the transmission delay between vertices. 
A synchronized solution is one where the states of all vertices are
identical,
\begin{equation}
x_{i}(t)=x(t)\quad\text{for all }i. \label{synchronization}%
\end{equation}
Thus $x(t)$ satisfies
\begin{equation}
x(t+1)=(1-\varepsilon)f(x(t))+\varepsilon f(x(t-\tau)). \label{sync-sol}%
\end{equation}

In order to investigate the stability of the synchronized solution (see
\cite{Pikovsky-book01} for a description of the stability concept employed here),
we consider orthonormal 
eigenmodes $u^k$ of the graph Laplacian
$\Delta_\Gamma$, defined by $(\Delta_\Gamma v)_i:=
(1/{n_i})\sum_{j \sim i}(v_j -v_i)$, 
with corresponding eigenvalues $-\lambda_k$.
The spectrum of $\Delta_\Gamma$ reflects the underlying connection topology,
with the zero eigenvalue $\lambda_0$ corresponding to the constant eigenfunction
\footnote{We follow here the conventions
of \cite{Jost02} where also further references can be found. The
eigenvalues are labelled in increasing order, the smallest one being
the trivial eigenvalue $\lambda_0=0$.}.
Since the
eigenfunctions yield an $L^2$-basis for functions on
$\Gamma$, it suffices to consider perturbations of $x$ of the form
\bel{3}
x_i(t)=x(t) + \delta \alpha_k(t) u^k_i(t)
\qe
for some small $\delta$, and for $k>0$, that is non-constant ones. 
The solution $x$
is stable against such a 
perturbation when $\alpha_k(t) \to 0$ for $t\to \infty$. Expanding
about $\delta=0$ yields
\begin{eqnarray} \label{4}
\alpha_k(t+1) &=& (1-\varepsilon) f'(x(t))\alpha_k(t) \nonumber \\
              &+& \varepsilon(1-\lambda_k) f'(x(t-\tau))\alpha_k(t-\tau).
\end{eqnarray}
The sufficient local stability condition is
\bel{5}
\lim_{T\to\infty} \frac{1}{T} \log\frac{|\alpha_k(T)|}{|\alpha_k(0)|} <0.
\qe
In the case without delay, that is, $\tau=0$, this is rewritten as
\bel{6}
 \lim_{T\to\infty} \frac{1}{T}
 \log\prod_{t=0}^{T-1}\frac{|\alpha_k(t+1)|}{|\alpha_k(t)|} <0 
\qe
and, by (\ref{4}), becomes
\bel{7}
\log|1-\varepsilon \lambda_k| + \lim_{T\to\infty} \frac{1}{T}\sum_{t=0}^{T-1}\log
|f'(x(t))| <0,
\qe
that is
\bel{8}
|e^{\mu}(1-\varepsilon \lambda_k)|<1
\qe
where $\mu=\lim_{T\to\infty} \frac{1}{T}\sum_{t=0}^{T-1}\log
|f'(x(t))|$ is the Lyapunov exponent of $f$.

Essentially the same reasoning works for the case of non-zero delay
$\tau$. Depending on whether $|\alpha_k(t)|$ is larger than
$|\alpha_k(t-\tau)|$ or not, in \rf{6}, we either keep the quotient
${|\alpha_k(t+1)}/{\alpha_k(t)|}$ or replace it by
${|\alpha_k(t+1)}/{\alpha_k(t-\tau)|}$. This reduces the number
of factors in \rf{6} by  $\tau$, but since we also get
correspondingly fewer factors in the term leading to the first term in
\rf{7}, the left hand side of \rf{7} is changed by a positive multiplicative
factor which does not affect the inequality. The terms get slightly
more complicated, as instead of 
\bel{9}
(1-\epsilon \lambda_k)f'(x(t)) \alpha_k(t)
\qe
yielding the two terms in \rf{7}, we now have 
\bel{10}
(1-\epsilon) f'(x(t))\alpha_k(t) + \varepsilon
(1-\lambda_k) f'(x(t-\tau))\alpha_k(t-\tau)
\qe
from \rf{4}. When comparing \rf{9} and \rf{10}, we see that in
\rf{10}, we may get additional partial cancellations due to different
signs of $f'(x(t))$ and $f'(x(t-\tau))$ or $\alpha_k(t)$  and
$\alpha_k(t-\tau)$, so that the absolute value can be significantly
smaller, thereby making the synchronization condition easier to
achieve. On the other hand, in a chaotic regime 
the values of $f'(x(t))$ and $f'(x(t-\tau))$ should be essentially
uncorrelated so that we still obtain the Lyapunov exponent $\mu$ as
the relevant parameter coming from $f$. 
A detailed analysis will appear elsewhere, but the foregoing ideas suggest that
synchronization in the delayed case is at least not more difficult,
but potentially easier (due to cancellations in \rf{10}) to achieve
than in the non-delayed one. Moreover, the connection topology, 
as reflected by the eigenvalues $\lambda_k$, still plays an important 
role in synchronization. 

\begin{figure}[tp]
\includegraphics[width=\mywidth, height=20cm]{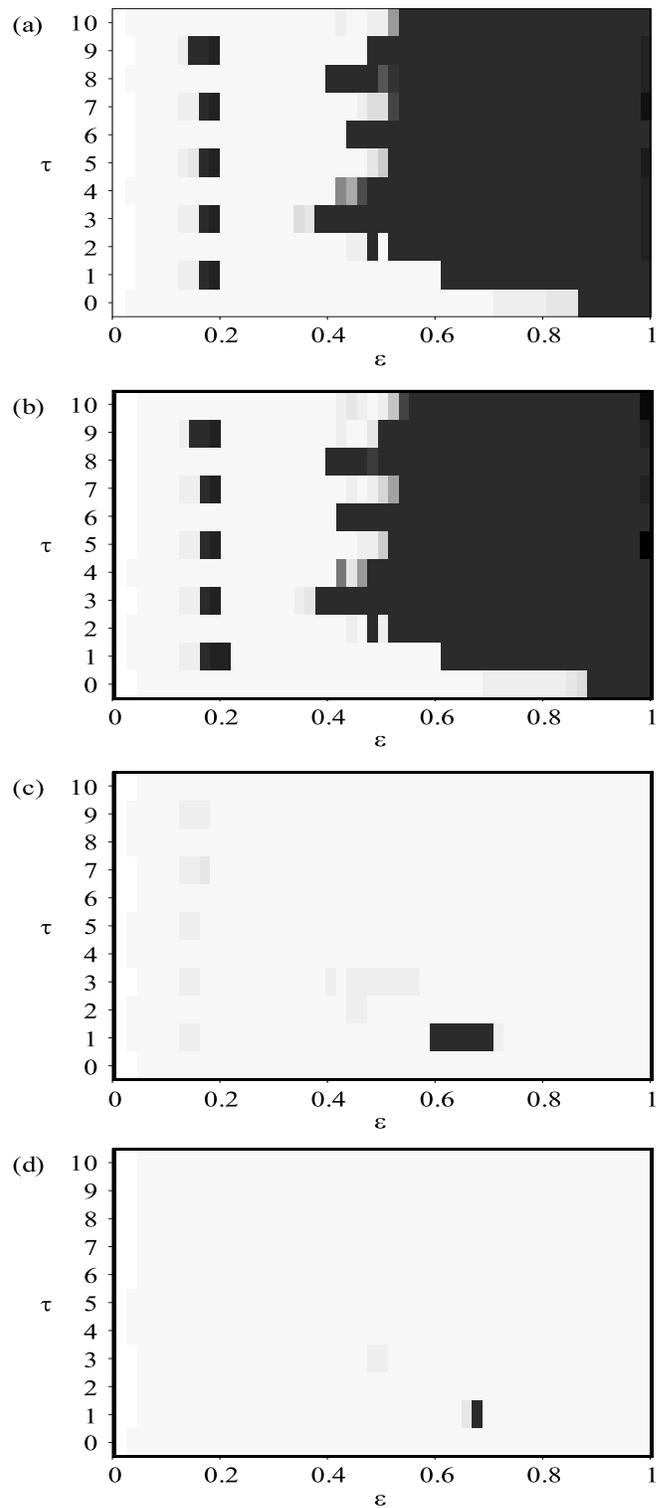}
\caption{Synchronization
of coupled logistic maps for different values of coupling strength
$\varepsilon$ and connection delays $\tau$, for the cases of (a) scale-free, (b)
random, (c) small-world, and (d) nearest-neighbor coupling. 
The grayscale encodes the degree of synchronization, 
with black regions corresponding to complete synchronization.}%
\label{fig:sync}%
\end{figure}

As an application, we take $f$ to be the logistic map%
\begin{equation}
f(x)=\rho x(1-x) \label{logistic}%
\end{equation}
which possesses a rich dynamical structure depending on the value of 
the parameter $\rho$ \cite{DiscreteChaos}.
Figure \ref{fig:sync} shows the
synchronization regions in the parameter space for several common network
architectures 
at the value $\rho=4$ for which $f$ is fully chaotic.
The gray-scale encoding represents the degree of synchronization
of the network after an initial transient of 10,000 time steps, as measured by
the fluctuations $\sigma^{2}(t)=\sum_{i=1}^{N}(x_{i}(t)-\bar{x}(t))^{2}$,
where 
$N$ is the size of the network and 
$\bar{x}(t)=(1/N)\sum_{i=1}^{N}x_{i}(t)$. 
Thus, $\sigma^{2}(t)\to 0$ as $t\to\infty$ if the system synchronizes.
Darker colors in the figure correspond to smaller values of $\sigma^2$, 
with black indicating that $\sigma^{2}(t)<10^{-25}$ after the
transients. 
The networks used in the simulations have the same size, $N=10,000$,
and the same number of average connections, even though the architectures may be different
\footnote{
The regular networks used in simulations have circular arrangement 
with each node coupled to its nearest $k$ neighbors, 
where $k$ is an even integer. 
Such networks are known to synchronize only for sufficiently small 
system sizes in the undelayed case \cite{Jost02}.
Small-world architecture is obtained by replacing randomly selected edges 
with long-range connections \cite{Watts98}.
Scale-free networks are constructed
using the algorithm of Barabasi and Albert \cite{Barabasi99},
starting with 10 isolated nodes and adding one node and 10 links at each step.
Each network consists of 10,000 nodes and about $10^5$ links.}.
The effects of the network topology are clearly seen in Figure~\ref{fig:sync}: 
Scale-free and random networks can synchronize for a large range of parameters whereas
more regular networks with nearest-neighbor and small-world type coupling do not.
In this respect, the similarities between scale-free and random networks 
(Figure \ref{fig:sync}a-b) are noteworthy. 
A closer inspection reveals some common features for synchronizing networks. 
For strong coupling (roughly for $\varepsilon>0.6$) synchronization is
achieved regardless of the actual value of the delay, as long as it is
positive. For intermediate coupling in the range $0.4<\varepsilon<0.6$, 
the value of the delay becomes decisive for synchronization.
In this range one may also observe long transients, clustering, 
and on-off intermittency \cite{Pikovsky-book01}, 
in the gray regions where $\varepsilon$
is slightly below the synchronization value.
There are also smaller regions of synchronization that exist 
for weaker coupling ($0.15<\varepsilon<0.20$) and only for odd delays.
Note that for zero delay synchronization can occur only for a
rather limited range ($\varepsilon>0.85$). 
Also, the small region of synchronization in Figure \ref{fig:sync}c-d   
occurs for nonzero delay. Therefore, the presence of
delays can indeed facilitate synchronization.

\begin{figure}[tb]
\fig{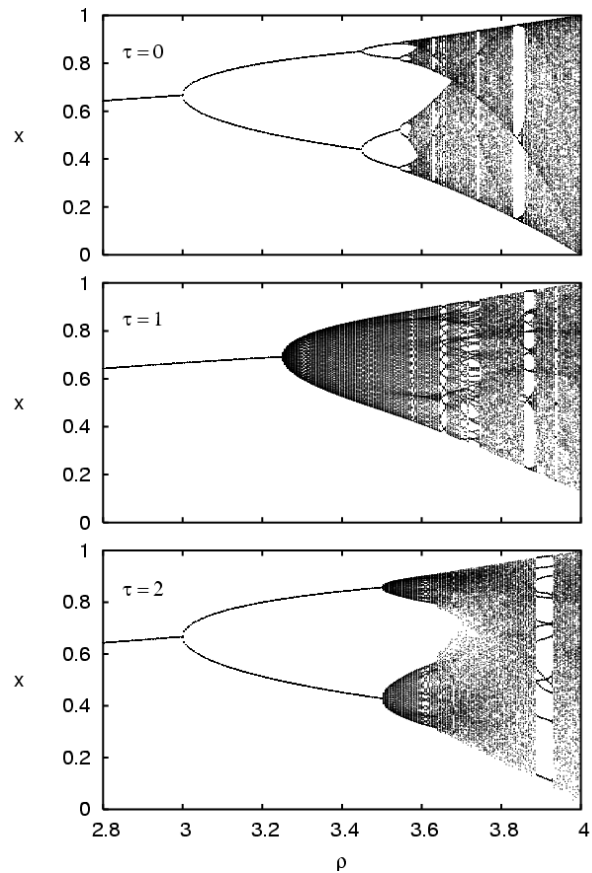} 
\caption{The
bifurcation diagram of the synchronized solution of coupled logistic maps for
several values of $\tau$. At each value of $\rho$, 200 iterates of
(\ref{sync-sol}) are plotted after inital transients. The value of
$\varepsilon$ is 0.8.}%
\label{fig:bif_rho}%
\end{figure}

\begin{figure}[t]
\fig{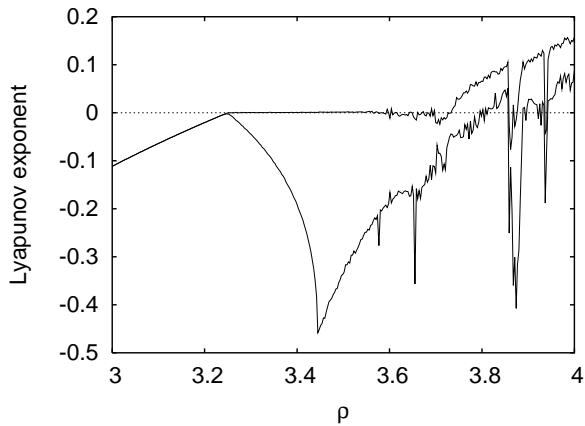}
\caption{The Lyapunov exponents of the synchronized solution for $\tau=1$
and $\varepsilon=0.8$.}%
\label{fig:lyap}%
\end{figure}

While the connection topology is important for the
synchronizability of the coupled system, 
connection delays are significant in determining 
the resulting synchronized dynamics. 
In particular, there is an important difference with the undelayed case, for
which (\ref{sync-sol}) reduces to
\begin{equation}
x(t+1)=f(x(t)); \label{map}%
\end{equation}
i.e., when the synchronized system can exhibit nothing different from the exact
dynamics of the individual isolated unit. Hence, an important implication of
connection delays is the possibility of the emergence of new collective
phenomena. 
Indeed,
the solutions of (\ref{sync-sol}) exhibit a much richer range of dynamics
when $\tau$ is nonzero. Figure~\ref{fig:bif_rho} shows the bifurcation diagram
for several values of $\tau$ when $f$ is given by
(\ref{logistic}). Of course, for $\tau=0$ the familiar bifurcation diagram of
the logistic map is obtained, displaying the period-doubling route to chaos.
By contrast, when $\tau>0$ Neimark-Sacker type bifurcations 
\cite{DiscreteChaos}
are prevalent,
immediately resulting in high-period solutions followed by more
complex behavior. 
Other values of odd and even delays result in pictures
similar to the cases $\tau=1$ and $\tau=2$, respectively. Since Neimark-Sacker
bifurcations cannot arise in one-dimensional maps, the difference with the
undelayed case is fundamental. 
Similarly, a consideration of the Lyapunov exponents shows that the character 
of the chaotic attractor is also different. 
For instance, as seen in Figure \ref{fig:lyap}, the synchronized system can 
have two positive Lyapunov exponents when connection delays are present, 
thus exhibiting hyperchaos. 

\begin{figure}[t]
\fig{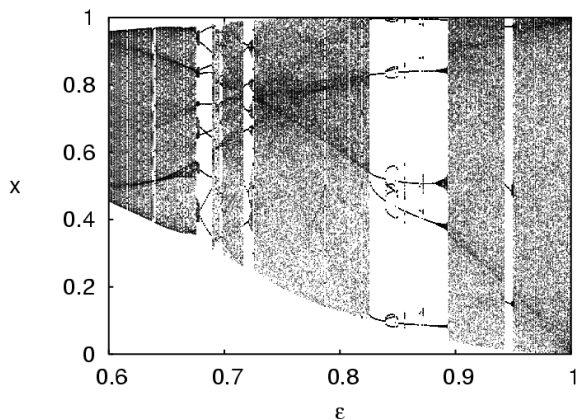}
\caption{The synchronized solution at different coupling strengths for
$\tau =1$ and $\rho=4$. 
At each value of $\varepsilon,$ 300 iterates of (\ref{sync-sol}) are
plotted after inital transients.}%
\label{fig:bif_eps}%
\end{figure}

\begin{figure}[t]
\fig{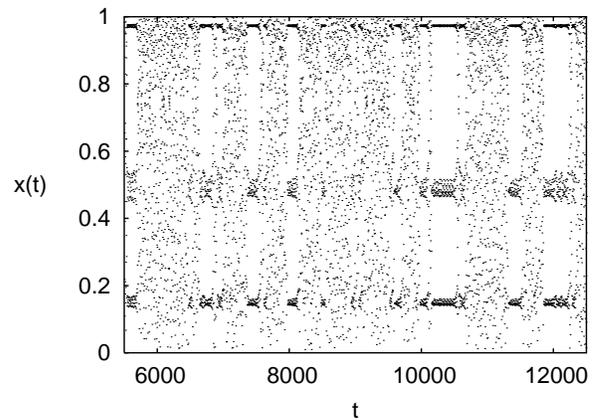}
\caption{Intermittent behavior of synchronized logistic maps for $\rho=4$, 
$\varepsilon=0.95$ and $\tau=1$. }%
\label{fig:intermittent}%
\end{figure}

Connection delays make the coupling strength $\varepsilon$ an important
factor in the dynamics of the synchronized solutions. To demonstrate,
we take the logistic map (\ref{logistic}) with $\rho=4$, for which it has a
chaotic attractor. Thus, (\ref{sync-sol}) is chaotic for all $\varepsilon$
when $\tau=0$. By contrast, when $\tau=1$ the attracting solutions of
(\ref{sync-sol}) display a wider variety depending on the value of
$\varepsilon$, as depicted in Figure~\ref{fig:bif_eps}. Chaos is interrupted
by windows of stable periodic solutions, for instance near $\varepsilon=0.68$
and $\varepsilon=0.71$. There is a large window of period-5 solutions for
$0.82<\varepsilon<0.89$, which also contain other stable solutions
of higher periods for certain $\varepsilon$ in this range.
A periodic attractor of period 3
exists near $\varepsilon=0.945$, which transforms to intermittent
behavior at $\varepsilon=0.95$, shown in Figure~\ref{fig:intermittent}. 
All these dynamics are exhibited by the coupled system (\ref{coupled}),
since Figure~\ref{fig:sync} implies that for these parameter values the system
is synchronized (provided it has an appropriate connection topology), so the
collective behavior is described by (\ref{sync-sol}).
Clearly, delays can make the dynamics of the synchronized system  
quite sensitive to the coupling strength,
a feature that is absent in undelayed networks. 

Networks with connection delays arise naturally in many areas of science,
including biology, population dynamics, neuroscience, economics, and so on. 
In neural systems, for instance, delays result from e.~g.~finite axonal transmission 
speeds \cite{Atay-Hutt-MPI03}. 
Such networks lack an intrinsic notion of simultaneity since the present 
state of the system is inaccessible to the constituent units. 
These networks are nevertheless capable of 
operating in synchrony, even when complex units and
significantly large delays are involved,
as our findings indicate.
In fact, networks with delays 
can actually synchronize more easily.
This synchronizability property is especially relevant for neural networks,
circumventing the difficulties in establishing a concept of collective
or simultaneous information
processing in the presence of delayed information transmission
\footnote{See also 
U.~Ernst, K.~Pawelzik, and T.~Geisel, Phys. Rev. Lett. \textbf{74}, 1570 (1995),
and W.~Gerstner, Phys. Rev. Lett. \textbf{76}, 1755 (1996),
in the context of coupled integrate-and-fire neurons.}.
Furthermore, the delays shape the dynamics of the synchronized system,
leading to the emergence of a variety of new dynamics which the individual units 
are not capable of producing 
\footnote{For further discussion along this line, see F.~M.~Atay and J.~Jost,
Max Planck Institute for Mathematics in the Sciences, 
Preprint Series Nr.~102 (2003), 
and arXiv:nlin.AO/0312026.}.
The observation that a wide range of different behavior
can be accessed by varying the coupling strength has important implications
for neural networks, where synaptic coupling strengths can change
through learning. 
This interesting connection provides additional motivation for investigating
the role of delays in complex adaptive systems.

\end{document}